\documentclass[11pt,twoside]{article}

\usepackage{asp2006}
\usepackage{epsf}
\usepackage{psfig}
\usepackage{lscape}

\begin{document}
\title{The 3D soft X-ray cluster-AGN cross-correlation function in the {\em ROSAT} NEP survey}   
\author{
N.~Cappelluti\altaffilmark{1},
H.~B\"ohringer\altaffilmark{1}
P.~Schuecker\altaffilmark{1},
E.~Pierpaoli\altaffilmark{2,6},
C.~R.~Mullis\altaffilmark{3},
I.~M.~Gioia\altaffilmark{4}, 
J.~P.~Henry\altaffilmark{1,5}
}
\altaffiltext{1}{Max Planck Institute f\"ur Extraterrestrische Physik, Postfach 1312, 85741, Garching, Germany}    
 \altaffiltext{2}{California Institute of Technology, Mail Code 130-33,Pasadena, CA 91125, USA }
\altaffiltext{3}{Department of Astronomy, University of Michigan, 918 Dennison, 500 Church Street, Ann Arbor, 48109-1042}
\altaffiltext{4}{Istituto di Radioastronomia INAF, Via P. Gobetti 101, 40129 Bologna, Italy} 
\altaffiltext{5}{Institute for Astronomy, University of Hawai'i, 2680 Woodlawn Drive, Honolulu, HI 96822  }
\altaffiltext{6}{ Physics and Astronomy Department, University of Southern California,  Los Angeles, CA 90089-0484, USA  }		  
\begin{abstract} 
X-ray surveys facilitate investigations of the environment
of AGNs. Deep {\textit Chandra} observations revealed that the AGNs
source surface density rises near clusters of galaxies.  The natural
extension of these works is the measurement of spatial clustering of
AGNs around clusters and the investigation of relative biasing between
active galactic nuclei and galaxies near clusters.The major aims of this work are to obtain a measurement of the
correlation length of AGNs around clusters and a measure of the
averaged clustering properties of a complete sample of AGNs in dense
environments. We present the first measurement of the soft X-ray cluster-AGN
cross-correlation function in redshift space using the data of the
{\em ROSAT}-NEP survey.  The survey covers $9\times9$ deg$^{2}$ around
the North Ecliptic Pole where 442 X-ray sources were detected and
almost completely spectroscopically identified.
We detected a $>$3$\sigma$ significant clustering signal on scales
$s$ $\leq$50 $h_{70}^{-1}$ Mpc. We performed a classical
maximum-likelihood power-law fit to the data and obtained a
correlation length $s_{0}$=8.7$^{+1.2}_{-0.3}$ $h_{70}^{-1}$ Mpc and a
slope $\gamma$=1.7$^{+0.2}_{-0.7}$ ( 1$\sigma$ errors).
This is a strong evidence that AGNs are good tracers of the large
scale structure of the Universe.  Our data were compared to the
results obtained by cross-correlating X-ray clusters and galaxies. We
observe, with a large uncertainty, that the bias factor of AGN is similar 
to that of galaxies. 
\end{abstract}
\vspace{-1cm}
\section{Introduction}
In the present paper, we concentrate on the study of the relative
clustering between X-ray selected AGNs and galaxy clusters. Our work
improves on most previous work on the large-scale structure of X-ray
selected AGNs in two important aspects. First, with the exception of
Mullis et al. (2004), our sample is the only one that is
spectroscopically complete (99.6\%). Gilli et al. (2005) used the CDFS
(35\%) and the CDFN (50\%). The Basilakos et al. (2005) sample had almost no
spectroscopic redshifts. Yang et al. (2006) used the CLASXS sample
(52\% complete) and the CDFN (56\% complete). Second, with the exception
of Mullis et al. (2004) and part of Yang et al. (2006), we measure a
three dimensional redshift space correlation function as opposed to
deprojecting the two dimensional angular correlation function.
 
 Another motivation for our work is that over the last several years,
  X-ray observations revealed that a
 significant fraction of high-$z$ clusters of galaxies show
 overdensities of AGNs in their outskirts (i.e. between 3 $h_{70}^{-1}$
Mpc and 7 $h_{70}^{-1}$ Mpc from the center of the cluster) (Henry
et~al., 1991; Cappi et~al., 2001; Ruderman \& Ebeling 2005, Cappelluti
et~al., 2005, and references therein). These overdensities were
however detected in randomly selected archive targeted observations of
galaxy clusters. While these overdensities are highly significant (up
to 8$\sigma$) when compared to cluster-free fields, the incompleteness
of the samples does not allow drawing any conclusion about the average
clustering properties of AGNs around clusters. The majority of the
sources making these overdensities have no spectroscopical
identification and therefore any information on their spatial
clustering is lost.  More recently Branchesi et al. (2007) showed that
at high-$z$ the source surface density of AGNs significantly increases
even in the central regions of the clusters.  These results imply that
further progress will come from studying the three dimensional spatial
distribution of AGNs around clusters.  A natural way to characterize
this specific type of clustering is given by the three-dimensional
{\it cross}-correlation of AGNs and galaxy clusters, the computation
of which needs complete redshift information for all objects, which is
rare in X-ray surveys.
\section{The Cluster AGN cross-correlation function}
The cross-correlation function $\xi_{CA}$ of clusters and AGNs is
defined by the joint probability to find, at a distance $r$, one
cluster in the infinitesimal comoving volume element $\delta V_{C}$
and one AGN in the comoving volume element $\delta V_{A}$,
$\delta~P=n_{C}n_{A}[1+\xi_{CA}(r)]\delta V_{C}\delta V_{A}$,
where $n_{C}$ and $n_{A}$ are the mean comoving number densities of
clusters and AGNs, respectively.  In calculating the differential
cross-correlation in redshift space we used an adapted version of the
Landy--Szalay estimator (Landy \& Szalay, 1993; see also e.g. Blake et
al., 2006).
\begin{figure}[!t]
\begin{center}
\hspace{0.cm}
\psfig{file=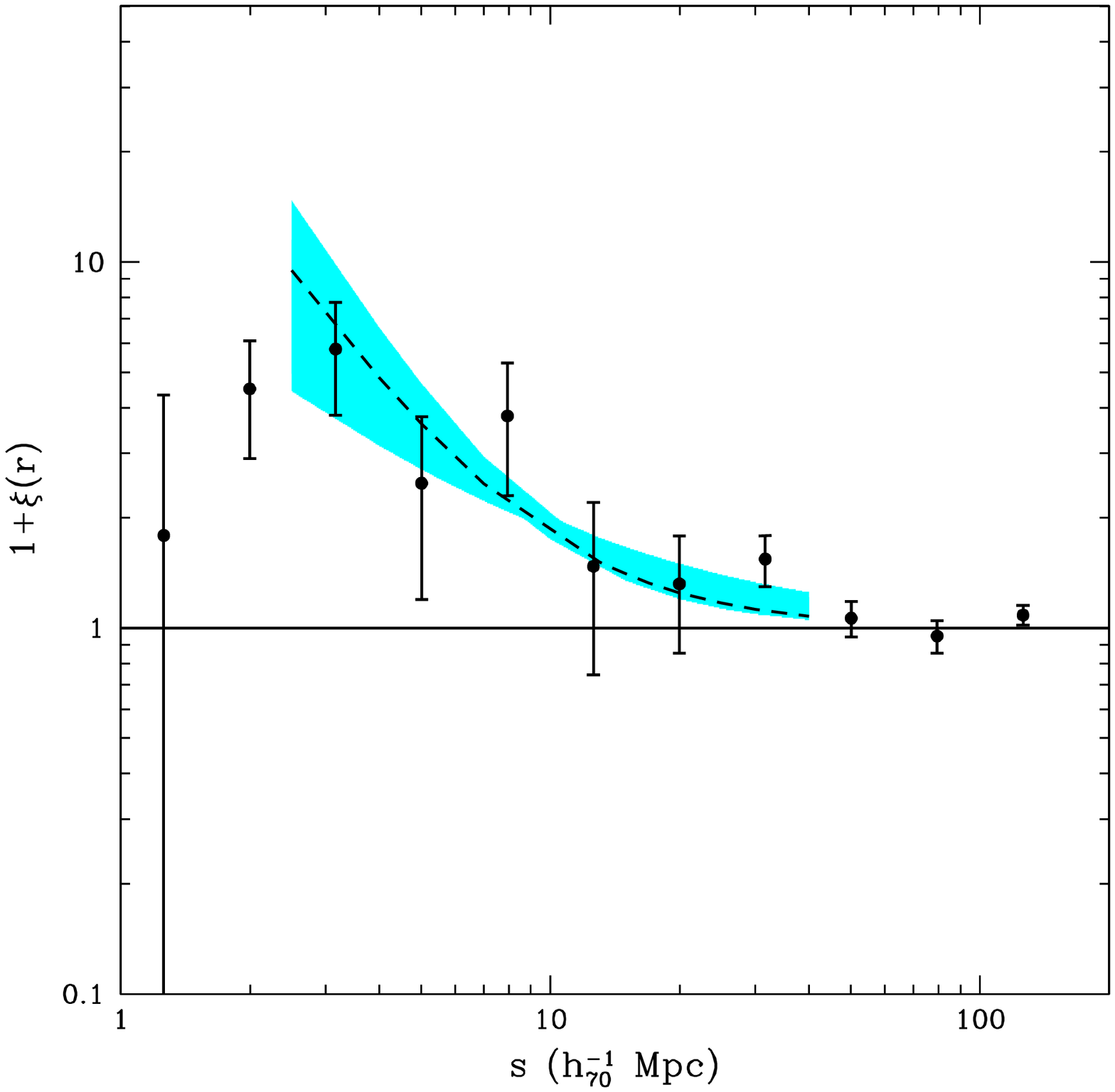,width=6.5cm,height=5.5cm,angle=0}
\psfig{file=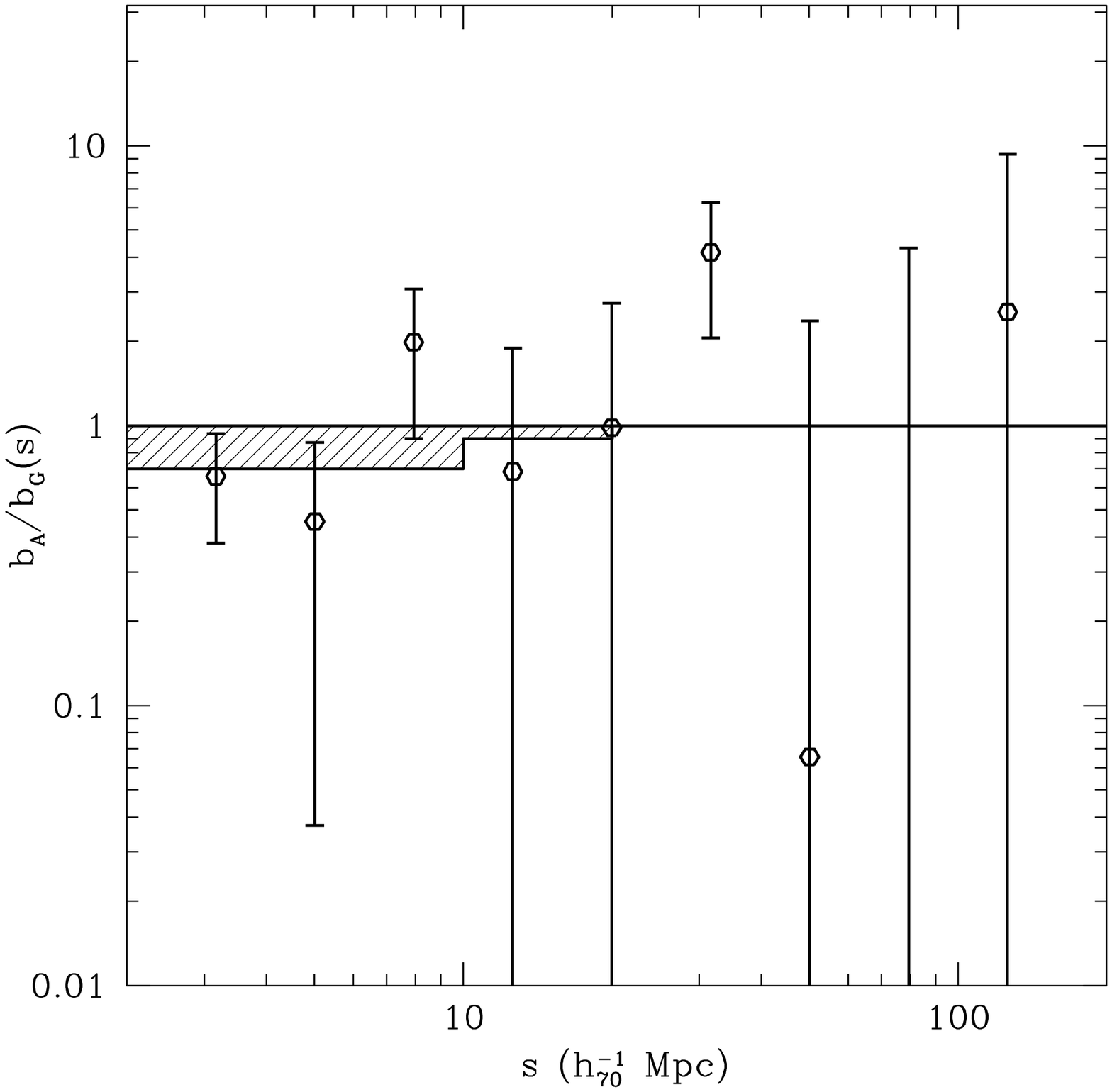,width=6.5cm,height=5.5cm,angle=0}
\caption{\label{fig:ccf}$Left~Panel:$ The Cluster-AGN soft X-ray cross correlation function plus one. 
The error bars are quoted at 1$\sigma$ level.  The dashed line represents the best fit 
maximum-likelihood power-law fit $s_{0}$=8.7$^{+1.2}_{-0.3}$ $h_{70}^{-1}$ Mpc and 
$\gamma$=1.7$^{+0.2}_{-0.7}$. The shaded region illustrates the 1$\sigma$ confidence 
region of the power-law fit in the distance range in which it was performed. 
$Right~Panel$: The ratio between the observed {\em ROSAT} NEP 
 $\xi_{CA}$(s) and the best fit  $\xi_{CG}$(r) obtained by S{\'a}nchez et al. (2005). 
 Errors  are quoted at the 1$\sigma$ level. The shaded region shows the expected level of 
$\frac{b_{A}}{b_{G}}(s)$=1 if the cross-correlation functions were compared in the  same space.}
\end{center}
\vspace{-1.cm}
\end{figure}

We present the spatial cross-correlation function between clusters  and AGNs in 
left panel of Fig. \ref{fig:ccf}. A positive clustering signal is detected  in the 
distance interval $s\leq$50 $h_{70}^{-1}$ Mpc. In order to test the strength of 
the clustering we performed a canonical power-law fit, $\xi_{CA}(s)=(\frac{s}{s_{0}})^{-\gamma}$,
with $s_{0}$  and $\gamma$  as  free parameters.
The best fit parameters obtained are 
$s_{0}$=8.7$^{+1.2}_{-0.3}$ $h_{70}^{-1}$ Mpc and 
$\gamma$=1.7$^{+0.2}_{-0.7}$  where the uncertainty is at the 1$\sigma$
confidence level. 
With $\gamma$ fixed to 1.8 (i.e. a typical value found in galaxy-galaxy correlation function)
 we find 
$s_{0}\sim 8.5 h_{70}^{-1}$ Mpc. A similar value was obtained by extending the fitting region
 to 60 $h_{70}^{-1}$ Mpc and restricting it to the 2.5--40 $h_{70}^{-1}$ Mpc.

\section{Discussion}

We presented here the first direct evidence of spatial clustering of soft X-ray 
selected AGNs around X-ray selected clusters of galaxies. Indirect evidence 
was presented by  Henry~et~al.~(1991), Cappi~et~al.~(2001), 
Cappelluti~et~al.~(2005) (and references therein).  These authors  found 
significant X-ray point source overdensities (about a factor 2) around distant
clusters of galaxies when compared to cluster-free fields. If the 
overdensities were at the cluster redshift they would arise at scales 
smaller than $\sim$7 $h_{70}^{-1}$ Mpc. Since the correlation function is proportional 
to $(\frac{\delta\rho}{\rho})^{2}$, a $\xi_{CA}$=1 implies an 
overdensity of a factor 2  with respect to a randomly distributed field. 
We can conclude that,  since the correlation length found in this work  
reflects the scale of the overdensities known up to now, we observe a physical 
overdensity (of at least a factor 2)  of AGNs around clusters between 
2 and $\sim$~8 $h_{70}^{-1}$ Mpc  from the center of the clusters.\\
Because of the shallowness of the NEP survey, the AGN surface density 
(i.e. $<$30 deg$^{-2}$ in the central region)
does not allow detection of such a correlation via overdensity analysis since it 
would be dominated by small number statistics. In fact, from our results we expect to detect 
AGNs overdensities on scales $<$7-8 $h_{70}^{-1}$ Mpc from the center of clusters. At $<z>\sim$0.18 (i.e. the 
median $z$ of the cluster sample of the NEP survey) these overdensities arise on scales
of $\sim$0.6 deg$^{-2}$ which are easily resolved by the  NEP survey.  
However to significantly detect these  overdensity on single clusters,
a conspicuous number of sources is necessary to disentangle real overdensities 
from  shot noise. 
As a final check we  
compared  our $\xi_{CA}$ to the X-ray cluster-galaxy   cross-correlation 
function (hereinafter CGCCF) computed by S{\'a}nchez et al. (2005). 
They  used the X-ray selected clusters of the REFLEX  survey (B\"ohringer et~al. 2002)
and the galaxies from the APM survey (Maddox et al. 1990) limited to b$_{j}$=20.5 mag. 
 They found that the CGCCF behaves like a broken power-law with a cut-off
distance of $\sim$2 $h_{70}^{-1}$ Mpc with a steeper slope at small distances. 
We can define the following approximate biasing relations:
\begin{equation}
\xi_{CA}(s)=b_{C} b_{A} \xi_{\rho}(s),\,\,\,\,\xi_{CG}(s)=b_{C}b_{G}\xi_{\rho}(s).
\end{equation}
Here $\xi_{\rho}(s)$ is the autocorrelation of matter, $b_{G},b_{A}$ and $b_{C}$
are the bias factors relative to  galaxies, AGNs and clusters, respectively.
By dividing the two equations we can then derive $\frac{b_{A}}{b_{G}}(s)$.
In order to perform this operation several effects must be taken in account.
The ratio $\frac{b_{A}}{b_{G}}(s)$  is plotted in right panel of Fig. \ref{fig:ccf} as a function of
the distance from the center of the cluster. The shaded region shows the value of our
measurement that implies that 
$\frac{b_{A}}{b_{G}}(s)$=1 when taking into account the difference between real and 
redshift space measurements discussed in the previous paragraph.
 The ratio is consistent with 1 on almost all  scales. We cannot exclude, 
within the errors, much different  values of the relative bias. Our data suggest an 
average relative bias consistent with unity
but allow an upper limit of $\sim$6  (at 1$\sigma$) at separations $s<50$ $h_{70}^{-1}$ Mpc. For separations
$s>10$ $h_{70}^{-1}$ Mpc no lower limits bigger than zero can be given. On 
larger scales the error increase  thus it is  difficult to draw any conclusion. At large separations
the power-law shape of $\xi_{CG}$ becomes uncertain, this  makes a comparison of our data
with those of  S\'anchez et~al.~(2005)  less meaningful.
We cannot exclude a significant antibiasing of AGNs when compared to galaxies, especially 
at low separations.
Though the amplitude of the uncertainties of our data still allows a fluctuation in the relative 
biasing of more than a factor 2, we can  conclude with a precision of 1$\sigma$ that the 
probability for a galaxy to become an  AGN is constant in the range of separations sampled in 
this work and that AGNs can be considered as good tracers of the dark matter distribution 
as are galaxies.

\end{document}